\documentclass[aps,reprint,superscriptaddress]{revtex4-1}
\usepackage{color}
\usepackage[version=3]{mhchem} 
\usepackage{gensymb}
\usepackage{amssymb}
\usepackage{upgreek}
\usepackage[usenames,dvipsnames]{xcolor}

\begin{document}

\title {Emergent formation of dynamic topographic patterns in electron beam induced etching}

\author{Aiden A. Martin}
\thanks{Equal contribution.}
\affiliation{Materials Technology for Energy Efficiency, University of Technology Sydney, Ultimo, New South Wales 2007, Australia}
\affiliation{Lawrence Livermore National Laboratory, Livermore, California 94550, USA}

\author{Alan Bahm}
\thanks{Equal contribution.}
\affiliation{Materials Technology for Energy Efficiency, University of Technology Sydney, Ultimo, New South Wales 2007, Australia}
\affiliation{FEI Company, 5350 Northeast Dawson Creek Drive, Hillsboro, Oregon 97124, USA}

\author{James Bishop}
\affiliation{Materials Technology for Energy Efficiency, University of Technology Sydney, Ultimo, New South Wales 2007, Australia}

\author{Igor Aharonovich}
\email{Igor.Aharonovich@uts.edu.au}
\affiliation{Materials Technology for Energy Efficiency, University of Technology Sydney, Ultimo, New South Wales 2007, Australia}

\author{Milos Toth}
\email{Milos.Toth@uts.edu.au}
\affiliation{Materials Technology for Energy Efficiency, University of Technology Sydney, Ultimo, New South Wales 2007, Australia}

\date{\today}

\begin{abstract}
\end{abstract}


\maketitle

{\bf 
Spontaneous formation of geometric patterns is a fascinating, ubiquitous process that provides fundamental insights into the roles of symmetry breaking, anisotropy and nonlinear interactions in emergent phenomena \cite{Cross1993851,Kondo20101616,Lehmann2014}. Here we report dynamic, highly ordered topographic patterns on the surface of diamond that span multiple length scales and have a symmetry controlled by the chemical species of a precursor gas used in electron beam induced etching (EBIE). This behavior reveals an underlying etch rate anisotropy and an electron energy transfer pathway that has been overlooked by existing EBIE theory. We present an etch rate kinetics model that fully explains our results and is universally applicable to EBIE. Our findings can be exploited for controlled wetting, optical structuring and other emerging applications that require nano and micro-scale surface texturing.
}

Electron beam induced etching (EBIE) \cite{Utke:2012ul,Toth2014} is a high resolution, single step, direct-write nanofabrication technique in which a precursor gas and an electron beam are used to realize etching. To date, EBIE has been used to machine a wide range of materials using etch precursor such as oxygen, water, ammonia, nitrogen trifluoride, xenon difluoride and chlorine. Key advantages of EBIE include cite-specificity and the ability to etch materials such as diamond which are resistant to conventional chemical etch processes. Consequently, EBIE has recently been used to realize practical device components for use in photonics \cite{Martin:2014km}, plasmonics \cite{Lassiter:2009kt} and nanofluidics \cite{Perry:2012fk}. 

In this work, we report an emergent pattern formation phenomenon caused by a chemical etch rate anisotropy in EBIE of single crystal diamond. The results reveal a shortcoming in existing, established EBIE theory which does not adequately explain the observed etch kinetics. We therefore propose a fundamental modification, whereby the critical role of energetic electrons is to transfer energy to and break bonds between surface atoms of the solid rather than to surface-adsorbed precursor molecules. The new EBIE model is confirmed experimentally, explains the observed patterns, and resolves long standing problems that have been identified in the EBIE literature. 

\begin{figure*}[h!]
\center
\resizebox{\textwidth*9/10}{!}{\includegraphics{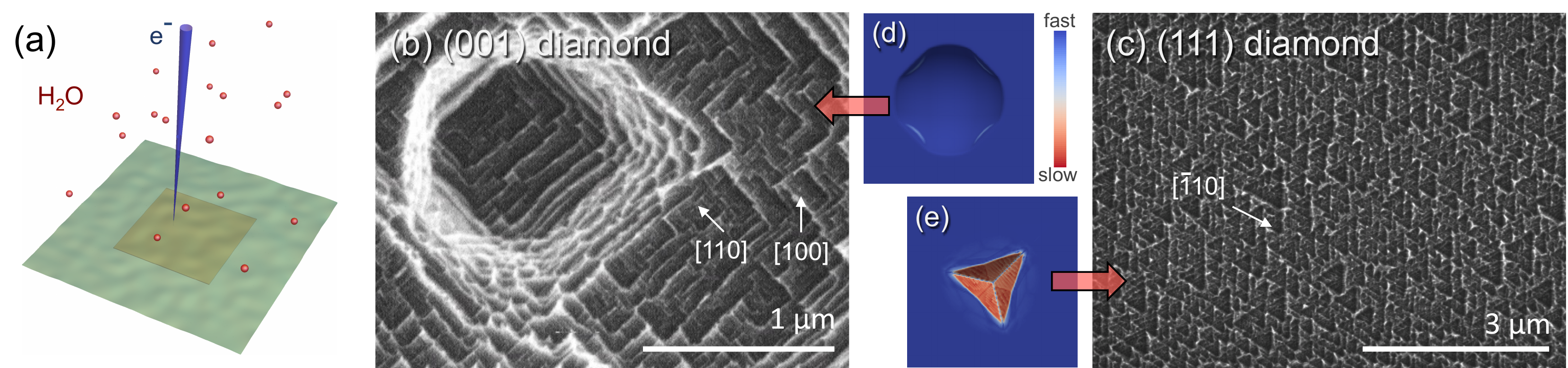}}
\caption{{\bf Topographic patterns formed during \ce{H2O} mediated electron beam induced etching of single crystal diamond.} (a) Schematic illustration of \ce{H2O} EBIE. (b) Expanding rhombohedra formed on the surface of (001) oriented diamond, and (c) trigons on the surface of (111) diamond. (d,e) Corresponding simulated rhombohedra and trigons (colored by the relative local etch rate) that are expected if the \{110\} planes are the slowest etching planes.}
\label{fig-patterns}
\end{figure*}

Figure \ref{fig-patterns}(a) is a schematic illustration of EBIE performed using \ce{H2O} precursor gas. Figure \ref{fig-patterns}(b) shows images of topographic patterns that form on the surface of single crystal (001) oriented diamond during \ce{H2O} EBIE performed at room temperature. A movie showing the pattern formation and evolution dynamics is provided as Supplementary Video \#1. Such patterns have not been reported previously because their formation is inhibited by small amounts of residual hydrocarbon contaminants present on the etched surface. Residual contaminants are very common in electron microscopes, alter the surface termination during EBIE and give rise to a competing process of electron beam induced deposition of carbon \cite{Lobo:2012rt}.  

Etching initiates at scratches and other surface defects, which expand laterally during EBIE, evolving into highly symmetric rhombohedra such as the large pit seen in the top left corner of Figure \ref{fig-patterns}(b). Similarly, the topography that is normally associated with surface roughness caused by EBIE rapidly evolves into step edges with \{110\} sidewalls which propagate laterally until they reach the edge of the area scanned by the electron beam. The \{111\} family of planes is absent from the resulting surface topography, the step sidewalls are comprised of \{110\} planes, and 90$\degree$ step corners are formed at the intercepts of the \{110\} planes. Corner formation requires the (110), ($\bar{1}$10), ($\bar{1}$$\bar{1}$0) and (1$\bar{1}$0) planes to etch slower than the (100), (010), (0$\bar{1}$0) and ($\bar{1}$00) planes. From these observations, we conclude that \ce{H2O}-mediated EBIE removes material from the \{100\} and \{111\} planes faster than from the \{110\} family of planes. 

In order to prove conclusively that the proposed anisotropy yields the four-fold symmetry observed in Figure \ref{fig-patterns}(b), we used the 3D implementation of the level set method (LSM) \cite{Osher:1608801}. LSM is a robust technique for evolving implicit surfaces under anisotropic velocity fields. It can be used to calculate surface shapes produced by etch rate anisotropies defined by differences between the etch rates of specific crystal planes. The simulations detailed in the Supporting Information reveal that the calculated surfaces match experiment only if the \{110\} planes that are oriented at $90\degree$ with respect to the electron beam axis etch slower than all other planes. Supplementary Video \#2 and Figure \ref{fig-patterns}(d) show the resulting surface features (rhombohedra). 

\begin{figure*}[h!]
\center
\resizebox{\textwidth*8/10}{!}{\includegraphics{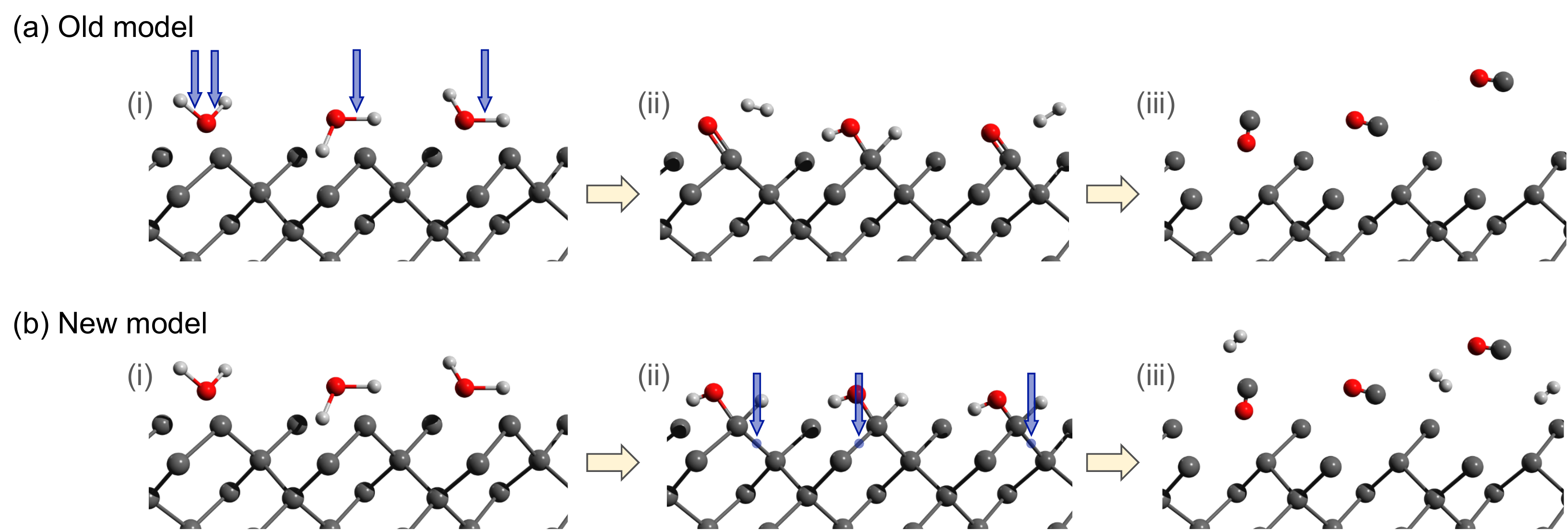}}
\caption{{\bf Simplified schematics of the mechanisms behind the standard and new EBIE models.} (a) Standard model: (i) Electrons dissociate physisorbed precursor molecules generating radicals that (ii) react with surface atoms and (iii) desorb spontaneously at the temperature used to perform EBIE. (b) New model: (i) Gas phase and physisorbed precursor molecules decompose spontaneously, and (ii) electrons break bonds in the crystal lattice causing (iii) the desorption of etch reaction products.}
\label{fig-EBIEmodel}
\end{figure*}

Validity of the simulation was confirmed by applying the same etch rate anisotropy rule set to (111) oriented diamond. The simulation and \ce{H2O} EBIE both produced the trigons shown in Figure \ref{fig-patterns}(c) and (e). The LSM simulations therefore support our conclusion that the geometries of patterns observed during \ce{H2O} EBIE of (001) and (111) oriented diamond is governed primarily by slow etching of specific \{110\} planes. This anisotropy can not be explained by conventional, established EBIE theory which is based on the mechanistic framework shown schematically in Figure \ref{fig-EBIEmodel}(a), where the key role of energetic electrons is to dissociate surface-adsorbed precursor molecules. In the case of \ce{H2O} EBIE of diamond, a possible pathway in this framework is the following \cite{Utke:2012ul,Toth2014}:
\begin{eqnarray}
\label{chem-ads}
\ce{H2O_{[v]}}~ & \leftrightarrow & ~\ce{H2O_{[p]}}\\
\label{chem-dissociate}
\ce{H2O_{[p]}}~ & \xrightarrow{\Xi_1} & ~\ce{H2} + \ce{O^*_{[c]}}\\
\label{chem-terminate}
\ce{O^*_{[c]}} + \ce{C_{[s]}}~ & \rightarrow & ~\ce{C_{[s]}O_{[c]}}\\
\label{chem-desorb}
\ce{C_{[s]}O_{[c]}}~ & \xrightarrow{\Xi_2} & ~\ce{CO_{[v]}}
\end{eqnarray}
where the subscripts [v], [s], [p] and [c] signify vapor phase, solid phase, physisorbed and chemisorbed species, respectively. $\Xi_1$ represents the energy barrier for dissociation of \ce{H2O}, and $\Xi_2$ is the binding energy of the reaction product. According to the standard EBIE model, $\Xi_1$ and $\Xi_2$ are overcome by a transfer of kinetic energy from the electrons that drive EBIE, and thermal energy of the substrate $(kT)$, respectively. This model has been used to explain a wide range of experiments such as dependencies of etch rates on time, beam current density and pressure of the precursor gas  \cite{toth2015,Toth2014,Perry:2012fk,cryo2014,Schoenaker:2011ko,Randolph:2011ia,Vanhove:2010fe,Miyazoe:2008ts,Lassiter:2008ul,Choi:2007vi,Rack:2003to,Taniguchi:1997uh,Fujioka:1990vg}. However, the model can not explain the etch rate anisotropy seen in Figure \ref{fig-patterns}, unless different crystal planes give rise to significant variations in the electron dissociation cross-section of \ce{H2O} adsorbates, the secondary electron emission yield, or the local coverage of precursor molecule adsorbates. None of these are plausible since the precursor is \ce{H2O}, the patterns form at room temperature, and the slowest etching planes are not consistently dark in secondary electron images.  

\begin{figure}[h!]
\center
\resizebox{\textwidth*5/11}{!}{\includegraphics{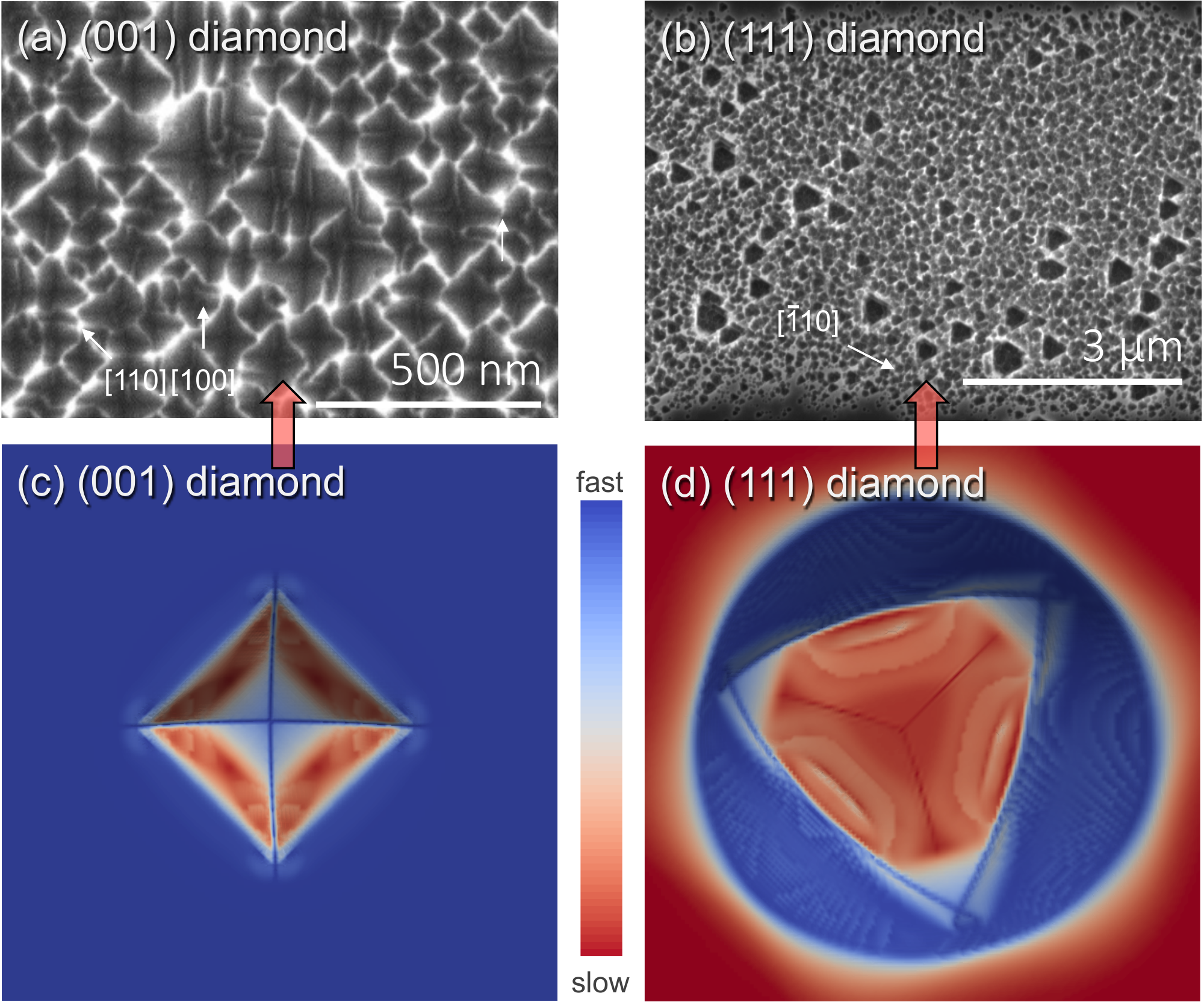}}
\caption{{\bf Topographic patterns formed during electron beam induced etching of single crystal diamond in the presence of \ce{NH3}.} (a) Expanding inverting pyramids formed on the surface of (001) oriented diamond, and (b) trigons on the surface of (111) diamond. (c,d) Corresponding simulated pyramids and trigons (colored by the relative local etch rate) that are expected if the \{111\} planes are the slowest etching planes.}
\label{fig-patternsNH3}
\end{figure}

To resolve the above issues, we propose a new mechanism, in which electrons provide the energy $\Xi_2$ in Reaction \ref{chem-desorb}, as shown in Figure \ref{fig-EBIEmodel}(b). That is, the critical role of electrons is not to dissociate the physisorbed precursor molecules, but to break bonds that bind surface atoms to the substrate and thus enable the desorption of the final reaction products. During etching, Reaction \ref{chem-dissociate} is expected to proceed spontaneously since active surface sites are generated continuously and the precursor molecules will dissociate on unterminated sites. In this framework, the etch rate anisotropy needed to produce the patterns seen in Figure \ref{fig-patterns} is expected since $\Xi_2$ (i.e. the C-C bond strength and the corresponding cross-section for scission by electrons) varies with the crystal plane. To confirm the proposed mechanism we performed an experiment based on the fact  \cite{Klauser:te} that the C-C bond strengths are modified by hydrogen which reconstructs and stabilizes the \{111\} surface. We therefore performed \ce{H2O} EBIE of (001) and (111) oriented diamond in the presence of \ce{NH3} gas, where the role of the \ce{NH3} is to supply an excess of hydrogen radicals to terminate the (111) planes. Figure \ref{fig-patternsNH3} shows that the corresponding surface patterns consist of inverted pyramids and trigons, respectively, and that these geometries are indeed expected from LSM simulations in which the \{111\} planes are the slowest etching planes.

\begin{figure}[h!]
\center
\resizebox{\textwidth*5/11}{!}{\includegraphics{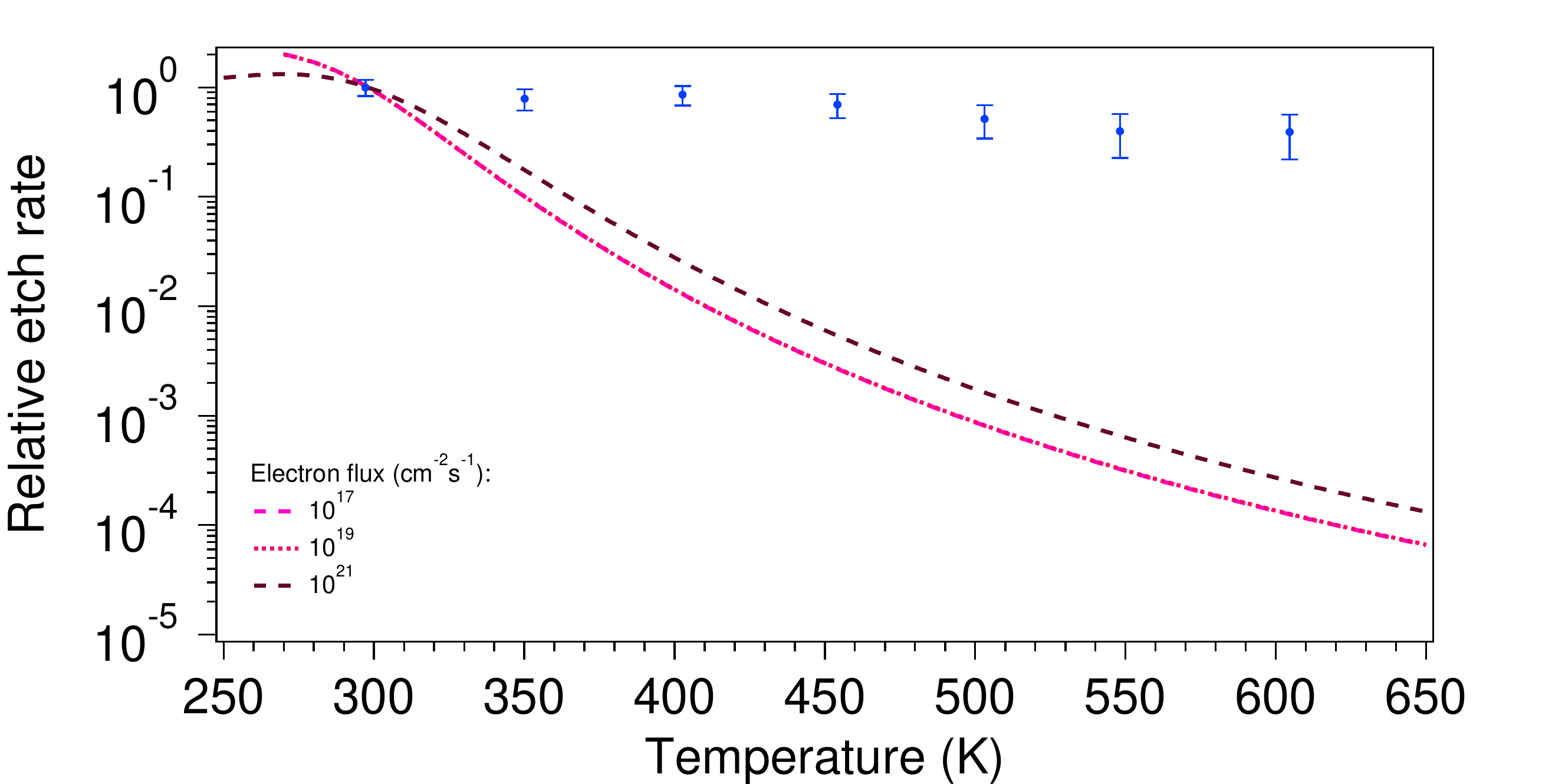}}
\caption{{\bf Temperature dependence of the rate of EBIE.} Rate of EBIE measured experimentally (points), and calculated using the established model of EBIE detailed in the Supporting Information (lines) using a wide range of electron fluxes. The etch rates are normalized to the rate of EBIE at room temperature.}
\label{fig-T}
\end{figure}

We performed one more experiment to further test the proposed EBIE mechanism. A consequence of the conventional EBIE model is that the EBIE rate is directly proportional to the concentration of physisorbed precursor molecules \cite{Utke:2012ul,Toth2014}. Hence, the etch rate of diamond is expected to be negligible at a temperature of $\sim 400$~K, as is shown in Figure \ref{fig-T}  (solid curves), irrespective of the electron flux used to perform EBIE. However, we observe significant etch rates at temperatures as high as 600~K. This result can not be explained by the conventional model, but is consistent with the new model in which the EBIE rate is proportional to the concentration of chemisorbed oxygen. The observed temperature dependence therefore serves as direct evidence for the new EBIE model. 

We note that the new EBIE model (Figure \ref{fig-EBIEmodel}(b)) is consistent with all results in the literature that were explained successfully by the conventional EBIE model. First, both models predict the same dependence of EBIE rate on electron beam energy since the secondary electron yield has the same dependence on beam energy as the amount of energy that is deposited by the beam into the surface atoms of the substrate. Second, both models predict the existence of reaction rate limited and mass transport limited etching regimes (as is shown in the Supporting Information), which makes both models consistent with a large amount of experimental data available in the EBIE literature. However, the new model is unique in being consistent with reports of UV laser induced etching of diamond, that is believed to proceed through a two photon C-C bond scission mechanism \cite{Lehmann2014}. Furthermore, the new model provides a satisfactory explanation for the fact that single crystal diamond can be etched by EBIE in the first place. The energy barrier of Reaction \ref{chem-desorb} in diamond is known to be significant \cite{Frenklach1993} and therefore etching observed at room temperature, or any temperature below the onset of defect generation and graphitization cannot be accounted for in the standard model.

Finally, we note that the topographic patterns cannot be explained by an anisotropic sub-surface damage generation mechanism analogous to the graphitization pathways encountered in conventional dry and wet diamond etch processes \cite{Shpilman2010,Olivero2005}. First, the etch rate anisotropy was modified significantly by the presence of \ce{NH3} gas, which should not change the sub-surface damage generation rate. Second, prior studies of EBIE of single crystal diamond have failed to produce any evidence of damage by photoluminescence and Raman spectroscopy \cite{Martin:2014km,Taniguchi:1997uh,1367-2630-14-4-043024,Martin:2015cx}. Third, the generation rate of damage produced by a 5~keV electron beam scales with the local energy density deposited into the substrate throughout the electron interaction volume \cite{Martin:2013fk}. The damage generation rate is therefore isotropic, except for special cases where the electron beam axis is parallel to a channeling axis, which should produce a strong dependence of the patterns on sample tilt, which was not observed in our experiments. We therefore conclude that sub-surface damage generation does not play a role in the observed etching and pattern formation behavior.

To summarize, we showed several dynamic pattern formations on the surface of single crystal diamond. We proposed an amended model for the EBIE process that is based on interactions of electrons with the substrate rather than the precursor molecule adsorbates. The new model explains our results and rectifies long standing issues in the literature of EBIE. Our results can be leveraged to engineer surface patterns controlled by electron beam irradiation conditions.

\section{Acknowledgements}

A portion of this work was funded by FEI Company and the Australian Research Council (Project Number DP140102721). A portion of this work was performed under the auspices of the U.S. Department of Energy by Lawrence Livermore National Laboratory under Contract DE-AC52-07NA27344. A.A.M. is the recipient of a John Stocker Postgraduate Scholarship from the Science and Industry Endowment Fund. I.A. is the recipient of an Australian Research Council Discovery Early Career Research Award (Project Number DE130100592). A.B. is grateful to Branislav Radjenovic and Ian Mitchell for insightful discussions of etch rate interpolation techniques and the level set method.

\bibliography{Bib-Patterns}

\begin{thebibliography}{10}
\expandafter\ifx\csname url\endcsname\relax
  \def\url#1{\texttt{#1}}\fi
\expandafter\ifx\csname urlprefix\endcsname\relax\def\urlprefix{URL }\fi
\providecommand{\bibinfo}[2]{#2}
\providecommand{\eprint}[2][]{\url{#2}}

\bibitem{Cross1993851}
\bibinfo{author}{Cross, M.~C.} \& \bibinfo{author}{{Hohenberg, P.C.}}
\newblock \bibinfo{title}{{Pattern formation outside of equilibrium}}.
\newblock \emph{\bibinfo{journal}{Rev. Mod. Phys.}}
  \textbf{\bibinfo{volume}{65}}, \bibinfo{pages}{851--1112}
  (\bibinfo{year}{1993}).

\bibitem{Kondo20101616}
\bibinfo{author}{Kondo, S.} \& \bibinfo{author}{Miura, T.}
\newblock \bibinfo{title}{{Reaction-diffusion model as a framework for
  understanding biological pattern formation}}.
\newblock \emph{\bibinfo{journal}{Science}} \textbf{\bibinfo{volume}{329}},
  \bibinfo{pages}{1616--1620} (\bibinfo{year}{2010}).

\bibitem{Lehmann2014}
\bibinfo{author}{Lehmann, A.}, \bibinfo{author}{Bradac, C.} \&
  \bibinfo{author}{Mildren, R.~P.}
\newblock \bibinfo{title}{{Two-photon polarization-selective etching of
  emergent nano-structures on diamond surfaces}}.
\newblock \emph{\bibinfo{journal}{Nat. Commun.}} \textbf{\bibinfo{volume}{5}},
  \bibinfo{pages}{3341} (\bibinfo{year}{2014}).

\bibitem{Utke:2012ul}
\bibinfo{author}{Utke, I.}, \bibinfo{author}{Moshkalev, S.} \&
  \bibinfo{author}{Russell, P.}
\newblock \emph{\bibinfo{title}{{Nanofabrication Using Focused Ion and Electron
  Beams: Principles and Applications}}} (\bibinfo{publisher}{Oxford University
  Press, USA}, \bibinfo{year}{2012}).

\bibitem{Toth2014}
\bibinfo{author}{Toth, M.}
\newblock \bibinfo{title}{{Advances in gas-mediated electron beam-induced
  etching and related material processing techniques}}.
\newblock \emph{\bibinfo{journal}{Appl. Phys. A Mater. Sci. Process.}}
  \textbf{\bibinfo{volume}{117}}, \bibinfo{pages}{1623--1629}
  (\bibinfo{year}{2014}).

\bibitem{Martin:2014km}
\bibinfo{author}{Martin, A.~A.}, \bibinfo{author}{Toth, M.} \&
  \bibinfo{author}{Aharonovich, I.}
\newblock \bibinfo{title}{{Subtractive 3D printing of optically active diamond
  structures}}.
\newblock \emph{\bibinfo{journal}{Sci. Rep.}} \textbf{\bibinfo{volume}{4}},
  \bibinfo{pages}{5022} (\bibinfo{year}{2014}).

\bibitem{Lassiter:2009kt}
\bibinfo{author}{Lassiter, J.~B.}, \bibinfo{author}{Knight, M.~W.},
  \bibinfo{author}{Mirin, N.~A.} \& \bibinfo{author}{Halas, N.~J.}
\newblock \bibinfo{title}{{Reshaping the plasmonic properties of an individual
  nanoparticle}}.
\newblock \emph{\bibinfo{journal}{Nano Lett.}} \textbf{\bibinfo{volume}{9}},
  \bibinfo{pages}{4326--4332} (\bibinfo{year}{2009}).

\bibitem{Perry:2012fk}
\bibinfo{author}{Perry, J.~M.}, \bibinfo{author}{Harms, Z.~D.} \&
  \bibinfo{author}{Jacobson, S.~C.}
\newblock \bibinfo{title}{{3D nanofluidic channels shaped by
  electron-beam-induced etching}}.
\newblock \emph{\bibinfo{journal}{Small}} \textbf{\bibinfo{volume}{8}},
  \bibinfo{pages}{1521--1526} (\bibinfo{year}{2012}).

\bibitem{Lobo:2012rt}
\bibinfo{author}{Lobo, C.~J.}, \bibinfo{author}{Martin, A.},
  \bibinfo{author}{Phillips, M.~R.} \& \bibinfo{author}{Toth, M.}
\newblock \bibinfo{title}{{Electron beam induced chemical dry etching and
  imaging in gaseous \ce{NH3} environments}}.
\newblock \emph{\bibinfo{journal}{Nanotechnology}}
  \textbf{\bibinfo{volume}{23}}, \bibinfo{pages}{375302}
  (\bibinfo{year}{2012}).

\bibitem{Osher:1608801}
\bibinfo{author}{Osher, S.} \& \bibinfo{author}{Fedkiw, R.}
\newblock \emph{\bibinfo{title}{{Level Set Methods and Dynamic Implicit
  Surfaces}}} (\bibinfo{publisher}{Springer}, \bibinfo{address}{New York},
  \bibinfo{year}{2003}).

\bibitem{toth2015}
\bibinfo{author}{Toth, M.}, \bibinfo{author}{Lobo, C.~J.},
  \bibinfo{author}{Friedli, V.}, \bibinfo{author}{Szkudlarek, A.} \&
  \bibinfo{author}{Utke, I.}
\newblock \bibinfo{title}{{Continuum models of focused electron beam induced
  processing}}.
\newblock \emph{\bibinfo{journal}{Beilstein J. Nanotechnol.}}
  \textbf{\bibinfo{volume}{6}}, \bibinfo{pages}{1518--1540}
  (\bibinfo{year}{2015}).

\bibitem{cryo2014}
\bibinfo{author}{Martin, A.~A.} \& \bibinfo{author}{Toth, M.}
\newblock \bibinfo{title}{{Cryogenic electron beam induced chemical etching}}.
\newblock \emph{\bibinfo{journal}{ACS Appl. Mater. Interfaces}}
  \textbf{\bibinfo{volume}{6}}, \bibinfo{pages}{18457--18460}
  (\bibinfo{year}{2014}).

\bibitem{Schoenaker:2011ko}
\bibinfo{author}{Schoenaker, F.~J.} \emph{et~al.}
\newblock \bibinfo{title}{{Focused electron beam induced etching of titanium
  with \ce{XeF2}}}.
\newblock \emph{\bibinfo{journal}{Nanotechnology}}
  \textbf{\bibinfo{volume}{22}}, \bibinfo{pages}{265304}
  (\bibinfo{year}{2011}).

\bibitem{Randolph:2011ia}
\bibinfo{author}{Randolph, S.}, \bibinfo{author}{Toth, M.},
  \bibinfo{author}{Cullen, J.}, \bibinfo{author}{Chandler, C.} \&
  \bibinfo{author}{Lobo, C.}
\newblock \bibinfo{title}{{Kinetics of gas mediated electron beam induced
  etching}}.
\newblock \emph{\bibinfo{journal}{Appl. Phys. Lett.}}
  \textbf{\bibinfo{volume}{99}}, \bibinfo{pages}{213103}
  (\bibinfo{year}{2011}).

\bibitem{Vanhove:2010fe}
\bibinfo{author}{Vanhove, N.}, \bibinfo{author}{Lievens, P.} \&
  \bibinfo{author}{Vandervorst, W.}
\newblock \bibinfo{title}{{Electron beam induced etching of silicon with
  \ce{SF6}}}.
\newblock \emph{\bibinfo{journal}{J. Vac. Sci. Technol. B}}
  \textbf{\bibinfo{volume}{28}}, \bibinfo{pages}{1206--1209}
  (\bibinfo{year}{2010}).

\bibitem{Miyazoe:2008ts}
\bibinfo{author}{Miyazoe, H.}, \bibinfo{author}{Utke, I.},
  \bibinfo{author}{Michler, J.} \& \bibinfo{author}{Terashima, K.}
\newblock \bibinfo{title}{{Controlled focused electron beam-induced etching for
  the fabrication of sub-beam-size nanoholes}}.
\newblock \emph{\bibinfo{journal}{Appl. Phys. Lett.}}
  \textbf{\bibinfo{volume}{92}}, \bibinfo{pages}{043124}
  (\bibinfo{year}{2008}).

\bibitem{Lassiter:2008ul}
\bibinfo{author}{Lassiter, M.~G.} \& \bibinfo{author}{Rack, P.~D.}
\newblock \bibinfo{title}{{Nanoscale electron beam induced etching: a continuum
  model that correlates the etch profile to the experimental parameters}}.
\newblock \emph{\bibinfo{journal}{Nanotechnology}}
  \textbf{\bibinfo{volume}{19}}, \bibinfo{pages}{455306}
  (\bibinfo{year}{2008}).

\bibitem{Choi:2007vi}
\bibinfo{author}{Choi, Y.~R.}, \bibinfo{author}{Rack, P.~D.},
  \bibinfo{author}{Frost, B.} \& \bibinfo{author}{Joy, D.~C.}
\newblock \bibinfo{title}{{Effect of electron beam-induced deposition and
  etching under bias}}.
\newblock \emph{\bibinfo{journal}{Scanning}} \textbf{\bibinfo{volume}{29}},
  \bibinfo{pages}{171--176} (\bibinfo{year}{2007}).

\bibitem{Rack:2003to}
\bibinfo{author}{Rack, P.~D.} \emph{et~al.}
\newblock \bibinfo{title}{{Nanoscale electron-beam-stimulated processing}}.
\newblock \emph{\bibinfo{journal}{Appl. Phys. Lett.}}
  \textbf{\bibinfo{volume}{82}}, \bibinfo{pages}{2326--2328}
  (\bibinfo{year}{2003}).

\bibitem{Taniguchi:1997uh}
\bibinfo{author}{Taniguchi, J.} \emph{et~al.}
\newblock \bibinfo{title}{{Electron beam assisted chemical etching of
  single-crystal diamond substrates with hydrogen gas}}.
\newblock \emph{\bibinfo{journal}{Jpn. J. Appl. Phys.}}
  \textbf{\bibinfo{volume}{36}}, \bibinfo{pages}{7691--7695}
  (\bibinfo{year}{1997}).

\bibitem{Fujioka:1990vg}
\bibinfo{author}{Fujioka, H.} \emph{et~al.}
\newblock \bibinfo{title}{{Measurements of the energy-dependence of
  electron-beam assisted etching of, and deposition on, silica}}.
\newblock \emph{\bibinfo{journal}{J. Phys. D}} \textbf{\bibinfo{volume}{23}},
  \bibinfo{pages}{266--268} (\bibinfo{year}{1990}).

\bibitem{Klauser:te}
\bibinfo{author}{Klauser, R.} \emph{et~al.}
\newblock \bibinfo{title}{{The interaction of oxygen and hydrogen on a diamond
  C(111) surface: a synchrotron radiation photoemission, LEED and AES study}}.
\newblock \emph{\bibinfo{journal}{Surf. Sci.}} \textbf{\bibinfo{volume}{356}},
  \bibinfo{pages}{L410--L416} (\bibinfo{year}{1996}).

\bibitem{Frenklach1993}
\bibinfo{author}{Frenklach, M.}, \bibinfo{author}{Huang, D.},
  \bibinfo{author}{Thomas, R.~E.}, \bibinfo{author}{Rudder, R.~A.} \&
  \bibinfo{author}{Markunas, R.~J.}
\newblock \bibinfo{title}{{Activation energy and mechanism of CO desorption
  from (100) diamond surface}}.
\newblock \emph{\bibinfo{journal}{Appl. Phys. Lett.}}
  \textbf{\bibinfo{volume}{63}}, \bibinfo{pages}{3090--3092}
  (\bibinfo{year}{1993}).

\bibitem{Shpilman2010}
\bibinfo{author}{Shpilman, Z.} \emph{et~al.}
\newblock \bibinfo{title}{{Oxidation and etching of CVD diamond by thermal and
  hyperthermal atomic oxygen}}.
\newblock \emph{\bibinfo{journal}{J. Phys. Chem. C}}
  \textbf{\bibinfo{volume}{114}}, \bibinfo{pages}{18996--19003}
  (\bibinfo{year}{2010}).

\bibitem{Olivero2005}
\bibinfo{author}{Olivero, P.} \emph{et~al.}
\newblock \bibinfo{title}{{Ion-beam-assisted lift-off technique for
  three-dimensional micromachining of freestanding single-crystal diamond}}.
\newblock \emph{\bibinfo{journal}{Adv. Mater.}} \textbf{\bibinfo{volume}{17}},
  \bibinfo{pages}{2427--2430} (\bibinfo{year}{2005}).

\bibitem{1367-2630-14-4-043024}
\bibinfo{author}{Schwartz, J.}, \bibinfo{author}{Aloni, S.},
  \bibinfo{author}{Ogletree, D.~F.} \& \bibinfo{author}{Schenkel, T.}
\newblock \bibinfo{title}{Effects of low-energy electron irradiation on
  formation of nitrogen--vacancy centers in single-crystal diamond}.
\newblock \emph{\bibinfo{journal}{New J. Phys.}} \textbf{\bibinfo{volume}{14}},
  \bibinfo{pages}{043024} (\bibinfo{year}{2012}).

\bibitem{Martin:2015cx}
\bibinfo{author}{Martin, A.~A.}, \bibinfo{author}{Randolph, S.},
  \bibinfo{author}{Botman, A.}, \bibinfo{author}{Toth, M.} \&
  \bibinfo{author}{Aharonovich, I.}
\newblock \bibinfo{title}{{Maskless milling of diamond by a focused oxygen ion
  beam}}.
\newblock \emph{\bibinfo{journal}{Sci. Rep.}} \textbf{\bibinfo{volume}{5}},
  \bibinfo{pages}{8958--4} (\bibinfo{year}{2015}).

\bibitem{Martin:2013fk}
\bibinfo{author}{Martin, A.~A.}, \bibinfo{author}{Phillips, M.~R.} \&
  \bibinfo{author}{Toth, M.}
\newblock \bibinfo{title}{{Dynamic surface site activation: a rate limiting
  process in electron beam induced etching}}.
\newblock \emph{\bibinfo{journal}{ACS Appl. Mater. Interfaces}}
  \textbf{\bibinfo{volume}{5}}, \bibinfo{pages}{8002--8007}
  (\bibinfo{year}{2013}).

\end{thebibliography}


\end{document}